\documentclass[article, draft]{agujournal2019} 
\usepackage{url} 
\usepackage{lineno}
\usepackage[inline]{trackchanges} 
\usepackage{multicol} 
\usepackage{soul}
\usepackage{amsmath}
\usepackage{amssymb}
\usepackage{natbib}
\usepackage{siunitx}
\usepackage[bitstream-charter]{mathdesign}
\justifying
\journalname{Geophysical Research Letters}

\newcommand{\shape}{\ensuremath{\mathrm{s}_{22}}}
\newcommand{\sparsity}{\ensuremath{\mathrm{s}_{21}}}

\DeclareMathOperator{\J}{J}

\begin{document}

\addeditor{JS}
\addeditor{AL}

\newcommand{\jc}[1]{\textcolor{Red}{[JC: #1]}}
\newcommand{\jws}[1]{\textcolor{Blue}{#1}}
\newcommand{\ael}[1]{\textcolor{Green}{[AL: #1]}}

\title{Characterizing Ocean Flows with \\ the Scattering Transform}

\authors{J. W. Skinner$^{1, 2}$, A. Lawrence$^{2}$, J. Callies$^{1}$}

%

\affiliation{1}{California Institute of Technology, Pasadena, CA}
\affiliation{2}{Brandeis University, Waltham, MA}

\correspondingauthor{J. W. Skinner}{jskinner@caltech.edu}




\begin{keypoints}
  \item The scattering transform is a data analysis method that can extract and quantify the morphology and distribution of features in fields.
  \item It can distinguish ocean flows that have significant geometric differences in physical space but may have the same power spectra.
  \item This can be used to identify waves and different flavors of geophysical turbulence in satellite altimetry and other snapshot observations.
\end{keypoints}


\section*{Abstract} 

Upper-ocean flows are a multi-scale jigsaw puzzle of turbulence and waves.
Characterizing these flows is essential for understanding their role in redistributing heat, carbon, and nutrients, yet power spectral analysis cannot always distinguish between types of motion.
We show that the scattering transform (ST), a wavelet convolution method, can extract geometric information from flow fields, offering insights beyond the power spectrum.
The ST distinguishes balanced dynamics, internal waves, and types of turbulence---even when their power spectra are identical.
Applied to sea surface height (SSH) fields from ocean models, the ST differentiates regions with distinct underlying dynamics.
Our analysis offers a framework for interpreting SSH from satellite altimetry missions and for analyzing other spatial maps (e.g., from airborne and coastal radar).
More generally, the ST is an appealing way to characterize complex fluid motion in a variety of geophysical contexts.

\section*{Plain Language Summary} 

Ocean currents take a range of different shapes: swift narrow currents along temperature fronts, swirls akin to weather systems in the atmosphere (albeit much smaller), and wavy motions excited by winds and tides.
It is not always easy to distinguish between these types of currents in observations and numerical simulations, but how they transport heat, carbon, and nutrients in the ocean is very different.
While waves reversibly deform tracers, different types of turbulent motions irreversibly stir and mix them.
Here, we describe how a new analysis technique, called the scattering transform, can characterize the geometry of flow structures and thereby offer important clues toward the underlying dynamics of observed and simulated ocean currents.

\section{Introduction} 

The upper ocean is a complex jigsaw puzzle of intertwined flows spanning a wide range of spatial and temporal scales.
At the mesoscale (\qtyrange{100}{300}{\kilo\meter}), eddies and jets advect heat, salt, carbon, and other tracers, shaping the large-scale circulation and biogeochemistry \citep[e.g.,][]{wunch_2007, morrow_2012}.
At the submesoscale (\qtyrange{1}{100}{\kilo\meter}), fronts, filaments, and eddies drive vertical nutrient fluxes and carbon sequestration, which may be critical for sustaining marine ecosystems and regulating atmospheric ${\rm CO}_2$ \citep[e.g.,][]{klein_2009, mahadevan_2016, mcwilliams_2016, levy_2012, levy_2018, taylor&thompson_2023}.
Internal waves, generated primarily by winds and tides, propagate through this dynamic system of currents, redistributing energy, shearing stratified layers, and facilitating small-scale mixing that impacts global climate \citep[e.g.,][]{jochum_2013,whalen_2020}.
Disentangling these different flows is crucial for understanding their individual contributions to transport, predicting oceanic responses to climate change, and improving ocean models.
This remains a significant challenge, particularly for submesoscale flows, because spatially resolved observations often lack the temporal resolution needed to separate turbulence and waves.

The power spectrum is an important tool for characterizing different flow regimes and their variability \citep[e.g.,][among many others]{Ferrari&Wunsch_2010, callies&ferarri_2013, callies_2015, Chereskinetal_2019, lawrence_2022}.
Many ocean flows are practically indistinguishable in their power spectra, however, because they follow the same power law.
For example, a forward energy cascade in three-dimensional turbulence \citep{Kolmogorov_1941}, an inverse energy cascade in two-dimensional turbulence \citep{kraichnan_1967}, and a forward cascade of surface potential energy in surface quasi-geostrophic turbulence \citep{blumen_1978} all predict a $k^{-5/3}$ power law for the kinetic-energy spectrum ($k$ is the horizontal wavenumber). 
In practice, this is also often indistinguishable from the empirical \citet{Garrett&Munk_1972} model for the internal-wave field, which has a $k^{-2}$ roll-off.
Tools beyond the power spectrum are therefore needed to characterize the distinct morphology of these flows.
Strongly nonlinear dynamics organize turbulent upper-ocean flows into eddies and filaments \citep{batchelor_1953, reeves_1969, townsend_1976, gargett_1989, frisch_1995, thorpe_2007, callies_2015}, whose spatial localization and coherence is encoded in the phase of the Fourier coefficients, which are discarded by the power spectrum \citep{armi_1985, Oppenheim&Lim_1981}.

\begin{figure}
  \centering
  \includegraphics[width=\textwidth]{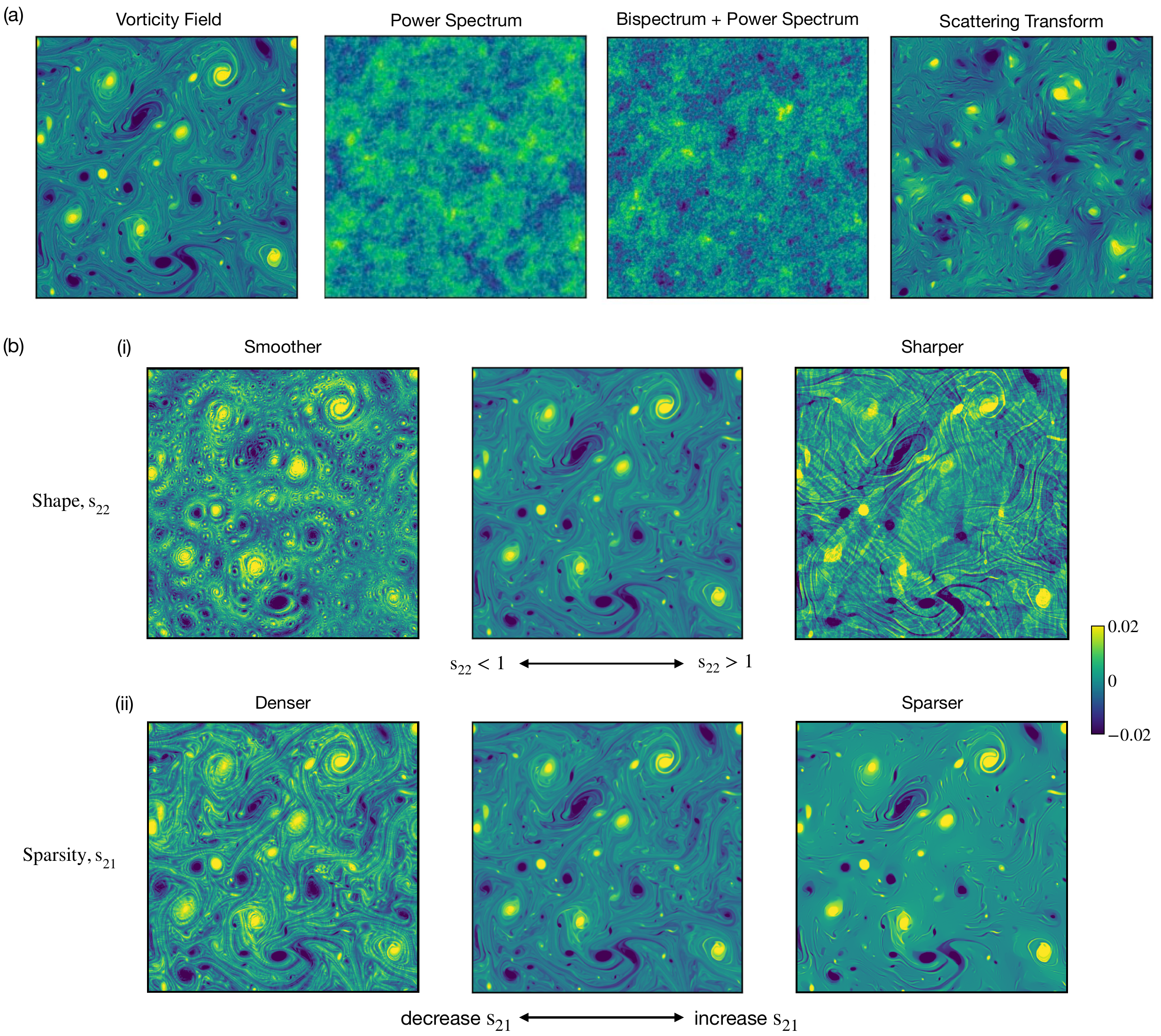}
  \caption{%
    Building intuition for the scattering transform.
    (a)~Comparison on structural information captured by the power spectrum, bispectrum, and scattering transform. 
    A turbulent vorticity field (left panel) with spatially intermittent and geometrically complex multi-scale structures is reconstructed using the information captured by each method.
    The power spectrum reconstruction randomizes the Fourier phases of the turbulent field; the bispectrum and scattering reconstructions optimize Gaussian random fields via gradient descent to match the respective statistics of the turbulent field \citep[following][]{cheng_2021}.
    (b)~Illustration of the diagnostics \shape\ (shape) and \sparsity\ (sparsity). 
    The turbulent field (middle panel) is modified to generate new fields where all \shape\ or \sparsity\ coefficients are modulated by $\pm 25 \%$.
 }
\label{fig:fig1}
\end{figure}

The scattering transform (ST) \citep{Mallat_2012} retains significantly more structural information than the power spectrum, making it effective for characterizing flow morphology (Fig.~\ref{fig:fig1}a).
While the bispectrum, which measures interactions between different wavenumber components, captures some structural information, it is still less effective than the ST \citep{Chengetal_2024}.
The ST captures geometric properties---e.g., localization, distribution, elongation, and orientation---through wavelet convolutions and modulus operations \citep{cheng_2021}.
The ST quantifies this information as a function of spatial scale, so dynamics at different scales can be probed.
In this letter, we apply the ST to idealized models that capture balanced flows, waves, and various turbulent regimes as well as to realistic ocean simulations that include internal gravity waves and meso- and submesoscale turbulence.
We specifically focus on characterizing flow structures in the sea surface height (SSH) field with an eye to the high-resolution spatial maps provided by the Surface Water and Ocean Topography (SWOT) mission \citep{Fuetal_2009, Morrowetal_2019, Fuetal_2024}.

\section{The Scattering Transform} 

The scattering transform (ST) is a representation of an input field $I = I(\mathbf{x})$, where $\mathbf{x} = (x, y)$, obtained by combining wavelet transforms, modulus operations, and spatial averaging \citep{Mallat_2012, bruna_2013}.
The first- and second-order scattering coefficients are
\begin{align}
    S_1(j_1, l_1) &= \langle |I \star \chi^{j_1, l_1}| \rangle, \label{eq:S1} \\
S_2(j_1, j_2, l_1, l_2) &= \langle ||I \star \chi^{j_1, l_1}| \star \chi^{j_2, l_2}| \rangle, \label{eq:S2}
\end{align}
where $\star$ denotes the convolution, \(\left| \, \cdot \, \right|\) the modulus, $\langle \, \cdot \, \rangle$ the spatial average, and $j_i$ and $l_i$ are the spatial scale $2^{j_i}$ (in pixels) and orientation of the wavelet~$\chi^{j_i, l_i}$ used in the $i$th convolution \citep{Mallat_2012, cheng_2020, cheng_2021}.
We generate each $\chi^{j, l}$ by dilating (indexed by $j$) and rotating (indexed by $l$) a Morlet wavelet, up to a maximum scale $J$ and with an orientation increment $2 \pi / L$.
The Morlet wavelet is defined by subtracting an offset $\beta$ from a plane wave and then localizing the result by modulating the wave by a Gaussian window: 
\begin{equation}
  \chi(\boldsymbol{x}) = \frac{1}{\sqrt{|\mathbf{\Sigma}|}} e^{-\boldsymbol{x}^\mathrm{T} \mathbf{\Sigma}^{-1} \boldsymbol{x} / 2} \left( e^{i \boldsymbol{k_0} \cdot \boldsymbol{x}} - \beta \right) \, .
\end{equation}
Here, $\beta = e^{-\boldsymbol{k}_0^\mathrm{T} \mathbf{\Sigma} \boldsymbol{k}_0/2}$; $\mathbf{\Sigma}$ is a covariance matrix that controls the Gaussian envelope's size, shape, and orientation; $\sigma = 0.8 \times 2^j$ is the width of the wavelet in real space (in pixels); and $\boldsymbol{k}_0$ is the wavenumber of the oscillation, with magnitude $|\boldsymbol{k}_0| = 3\pi /(4 \times 2^j)$ \citep[see][]{Andreuxetal_2020, cheng_2021, bernardino_2005}.  
The iterative convolution of the input field $I$ with $\chi^{j, l}$, followed by modulus operations, captures dependencies between wavenumber components of $I$ that preserve structural information that are lost by the power spectrum and the bispectrum \citep[see][and Fig.~\ref{fig:fig1}a]{armi_1985, Oppenheim&Lim_1981}.
While various wavelets, dilation factors, and orders can be used, we find that using a Morlet wavelet with a dyadic (scales increase by powers of two) dilation and computing scattering coefficients up to second order provides sufficient precision for distinguishing different ocean dynamics.

To efficiently interpret the ST information, we focus on two summary statistics that quantify the geometry of features in $I$ (\citealp{cheng_2021}; see also related work by \citealp{allys_2020}):
\begin{equation}
  {\rm s}_{21}(j_1, j_2) = \langle S_2 / S_1 \rangle _{l_1, l_2}, \qquad  {\rm s}_{22}(j_1, j_2) = \langle S_2^{\parallel} / S_2 ^{\perp} \rangle _{l_1}.
  \label{shape_sparsity}
\end{equation}
Here, $\langle \, \cdot \, \rangle_{l_i}$ denotes an average over orientations $l$, and $\parallel$ and $\perp$ denote orientations $l$ that are parallel and perpendicular to one another.

The meaning of \eqref{shape_sparsity} is illustrated by generating fields with increased or decreased values of ${\rm s}_{21}$ and ${\rm s}_{22}$ \citep[Fig~\ref{fig:fig1}b; see also][]{cheng_2021}.
We start with a vorticity snapshot (center panels) from a doubly periodic two-dimensional turbulence simulation and generate new fields with the same power spectrum but with ${\rm s}_{21}$ and ${\rm s}_{22}$ modulated in amplitude across all scales by $-25\%$ (left panels) and $+ 25\%$ (right panels).
The ${\rm s}_{21}$ quantify the ``sparsity'' (clustering) of features in $I$.
Higher ${\rm s}_{21}$ values indicate more localized structures in $I$.
Increasing ${\rm s}_{21}$ removes filaments, isolating vortices, while decreasing ${\rm s}_{21}$ adds filaments, filling the space between vortices.
The ${\rm s}_{22}$ quantify the ``shape'' of features in $I$. 
For ${\rm s}_{22} < 1$, structures are smooth or ``blobby'', and for ${\rm s}_{22} > 1$, structures are elongated and anisotropic.
Increasing \shape\ accentuates the filaments, whereas decreasing \shape\ enhances the vortices and smooths filaments.
The \sparsity\ and \shape\ values depend on the spatial scales $j_1$ and $j_2$, meaning that they capture structural information across different scales, providing a multiscale characterization of the flow field.

\section{Different Flavors of Turbulence} 

\begin{figure}
  \centering
  \resizebox{1.1\textwidth}{!}{\includegraphics[width=1.\textwidth]{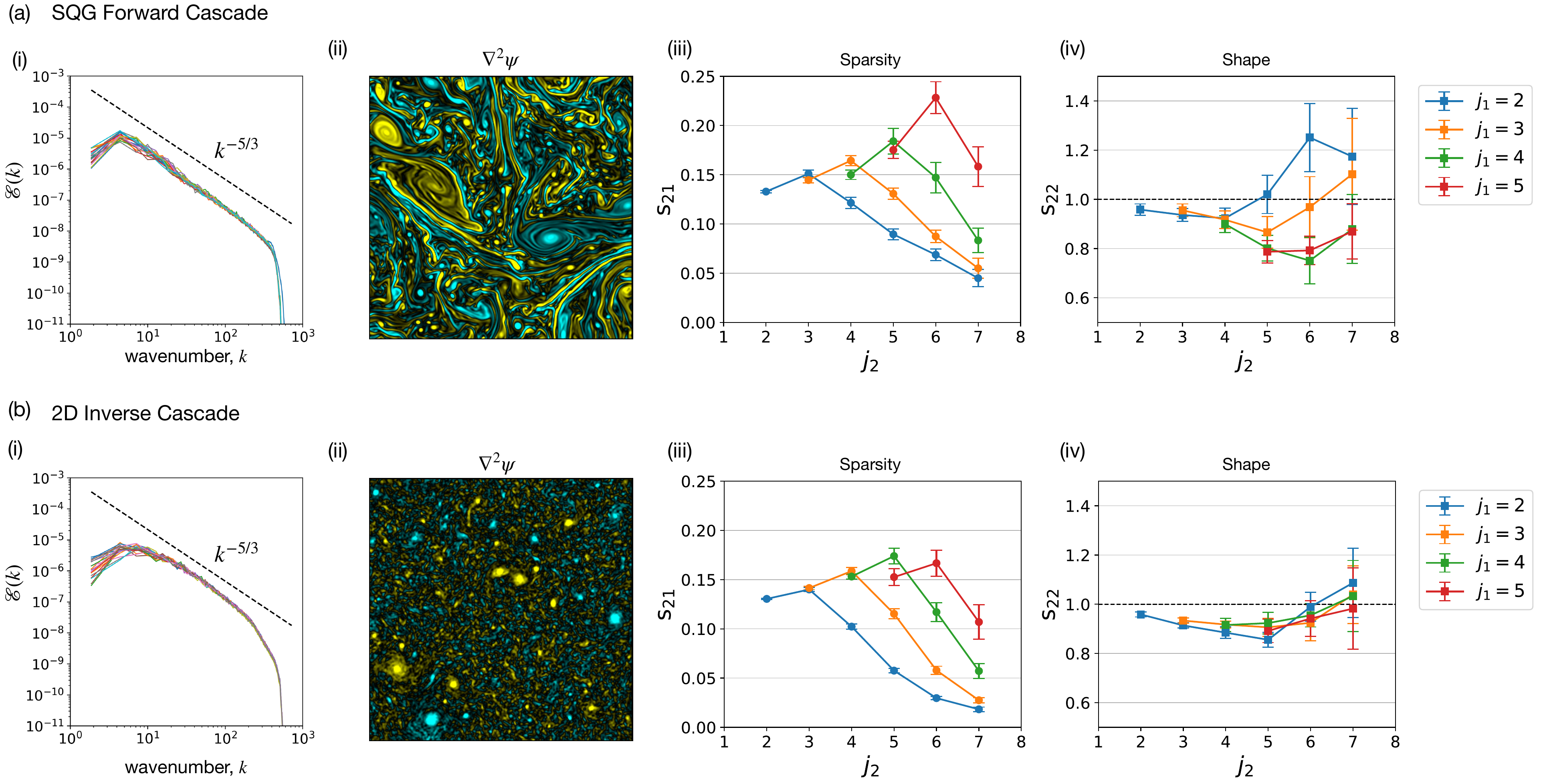}}
  \caption{%
  Distinguishing between two flavors of geophysical turbulence: (a)~SQG turbulence with a forward surface potential-energy cascade and (b)~2D turbulence with an inverse energy cascade.
  (i)~Kinetic-energy spectra of the 20 ensemble members.
  (ii)~Representative snapshots of the vorticity field $\nabla^2 \psi$.
  (iii,~iv)~Shape and sparsity coefficients averaged over ensemble members and their standard deviations (error bars).
  Shape and sparsity are shown for the spatial scales $j_1 = \numrange{2}{5}$ and $j_2 = \numrange{2}{7}$.
  These capture correlations between \numrange{4}{32} grid points and \numrange{4}{128} grid points in our $256 \times 256$ grid points domain.
  We exclude the grid and domain scales because they are affected by damping.
  In the shape plots, dashed lines indicate the transition from sharp ($\shape > 1$) to smooth ($\shape < 1$) structures. 
  }
\label{fig:fig2}
\end{figure}

The shape and sparsity coefficients \shape\ and~\sparsity\ can quantify structural differences between different flavors of geophysical turbulence that are indistinguishable by the power spectrum.
An inverse energy cascade in two-dimensional (2D) turbulence and a forward cascade of surface potential energy in surface quasi-geostrophic (SQG) turbulence both produce kinetic-energy spectra $\mathcal{E}(k) \propto k^{-5/3}$ \citep{kraichnan_1967, blumen_1978}, despite producing distinct flow structures dominated by isolated vortices in 2D and filaments with sharp edges interspersed by isolated vortices in SQG (Fig.~\ref{fig:fig2}).
We run 20-member ensembles of doubly periodic forced and dissipated simulations of 2D and SQG turbulence configured to equilibrate with identical spectral slopes in their inertial ranges (Fig.~\ref{fig:fig2}).
We vary the forcing wavenumber for each form of turbulence: the inertial ranges are above the forcing scale in 2D turbulence, which drives an inverse energy cascade from small to large scales, and below the forcing scale in SQG turbulence, which drives a forward surface potential-energy cascade from large to small scales \citep{blumen_1978, Pierrehumbert_1994, held_1995, capet_2008}.
In both, energy is removed at large scales to enable equilibration.
Under these conditions, 2D and SQG turbulence develop $\mathcal{E}(k)$ with the same power law (Fig.~\ref{fig:fig2} panels~i).
The markedly different flow structures produced by the different underlying turbulent dynamics (Fig.~\ref{fig:fig2} panels~ii) can be captured and quantified by the ST.

We use \shape\ and \sparsity\ to distinguish between 2D and SQG turbulence, applying the ST to the vorticity fields $\nabla^2 \psi$, where $\psi$ is the streamfunction.
For each ensemble member, we analyze a single snapshot of $\nabla^2 \psi$ taken at the same late time in the simulation, when the equilibrium has been reached.
Mean and standard deviations of \shape\ and \sparsity\ are calculated over the 20~ensemble members to distinguish between random fluctuations of these diagnostics and differences arising from the underlying characteristics of the turbulence.
Clear qualitative and quantitative differences in \shape\ and \sparsity\ occur across nearly all spatial scales (Fig.~\ref{fig:fig2} panels iii and~iv).

SQG turbulence is sparser than 2D turbulence, with \sparsity\ differences being more pronounced at larger scales ($j_2 \sim 5$).
Shape provides a particularly strong distinction between SQG and 2D turbulence across nearly all scales.
SQG turbulence is sharper (${\rm s}_{22} > 1$) across scales, while 2D turbulence is smooth (${\rm s}_{22} < 1$).
These values align with visual inspection of the $\nabla^2 \psi$ fields.
2D turbulence appears more homogeneous with smooth and evenly distributed vortices, while SQG turbulence has fewer larger vortices linked by sharp and densely packed filaments \citep[cf.,][]{held_1995}.
Essentially all \shape\ and \sparsity\ have standard deviations much smaller than the differences between the SQG and 2D cases, indicating that even single snapshots can contain sufficient information to robustly discriminate between the two turbulence regimes using the ST.

\section{Waves and Balanced Flow} \label{sec:sec4} 

Separating balanced flows from waves in sea surface height (SSH) fields is crucial for understanding ocean dynamics from satellite altimetry data \citep{Chavanne&Klein_2010, richman_2012, Arbicetal_2013, Torresetal_2018, callies&wu_2019}.
Balanced flows primarily drive transport along isopycnals \citep[e.g.,][]{marshall_2006}, whereas internal waves contribute to irreversible mixing through vertical shear instabilities and small-scale turbulence that crosses isopycnals \citep[e.g.,][]{garrett_kunze_2007}.
While waves and balanced flows are distinguishable by their different time scales, this is not practical for altimetry data because snapshots are too infrequent.
Velocity-based methods \citep{Buhleretal_2014} are also not applicable because the relationship between SSH gradients and velocities differs for waves and balanced flows.

Among internal waves, low-mode internal tides leave the strongest imprint on SSH, with distinct peaks at the modal wavenumbers corresponding to semidiurnal and diurnal tidal frequencies \citep[e.g.,][]{richman_2012, ray&zaron_2016, callies&wu_2019}.
Identifying internal tides in SSH data is complicated, however, because their amplitude is often comparable to that of balanced turbulence, and topographic effects can broaden their wavenumber signature \citep{buhler&muller_2007}.
While harmonic fits to time series are commonly used, they can miss variability caused by transient mesoscale activity \citep{rainville_propagation_2006} and seasonally changing stratification \citep{Gerkemaetal_2004}.
These factors can shift the phase and amplitude of internal tides over time, leading to a decrease in the variance captured by harmonic fits.
Moreover, SSH variability is not solely due to internal waves at tidal frequencies.
A broadband spectrum of non-tidal internal waves---generated by wind forcing, wave--wave interactions, spontaneous loss of balance, and other processes---generally overlaps with the other components of the signal in wavenumber space and complicates the interpretation of SSH spectra.

\begin{figure}
  \centering
  \includegraphics[width=1\textwidth]{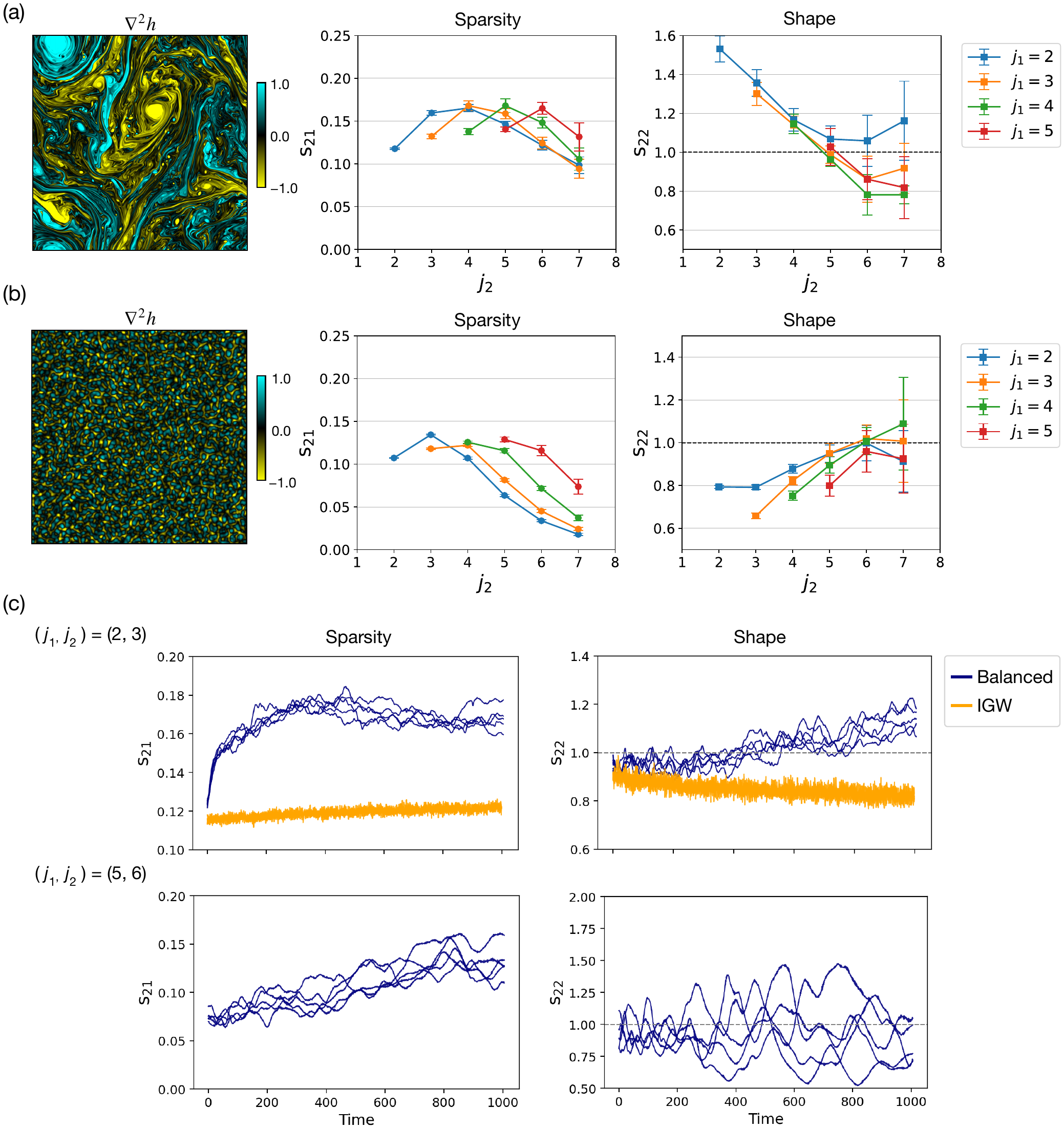}
    \caption{%
    Distinguishing balanced flows and inertia--gravity waves (IGWs).
    (a)~Balanced and (b)~IGW flow fields and their shape and sparsity from shallow-water simulations.
    Snapshots of $\nabla^2 h$, where $h$ is surface height, at $t = 1000$ are shown alongside shape and sparsity coefficients averaged over 20~ensemble members (error bars show standard deviations).
    (c)~Time series of shape and sparsity for individual simulations in each ensemble at spatial scales $j_1 = 2$, $j_2 = 3$ (\numlist{4;8} grid points) and $j_1 = 5$, $j_2 = 6$ (\numlist{32;64} grid points).
    The domain size is $256 \times 256$ grid points.
    Balanced and IGW simulations are indicated by navy and orange lines, respectively.
    }  
  \label{fig:fig3}
\end{figure}

We test the ST's ability to distinguish balanced turbulence and inertia-gravity waves (IGWs) in surface height using doubly periodic and freely decaying solutions to the shallow-water equation (SWE).
Simulations are initialized with either linear IGWs or geostrophic flow (balanced), both with a $\mathcal{E}(k) \propto k^{-2}$ kinetic-energy spectral slope and random phases.
The IGW field remains dominated by linear wave dynamics, with energy only gradually dispersing across scales by weak nonlinear interactions.
The strongly nonlinear balanced flow generates thin filaments and undergoes an inverse energy cascade via the merging of vortices to larger vortex structures.
Snapshots of $\nabla^2 h$ illustrate the differences between the IGW and balanced fields at the end of their evolutions (Fig.~\ref{fig:fig3}a,b).
We apply the ST to the $\nabla^2 h$ fields; and, as before, average \shape\ and \sparsity\ over 20 ensemble members for the balanced and IGW cases, each initialized with a different random seed.
We assess ensemble averaged \shape\ and \sparsity\ for scales $j_1=\numrange{2}{5}$ and $j_2=\numrange{2}{6}$ and calculate standard deviations over the ensembles.

IGW and balanced flows are clearly distinguished by the ST (Fig.~\ref{fig:fig3}a,b) and are also separated in the time series of their \shape\ and \sparsity\ (Fig.~\ref{fig:fig3}c).
While different flavors of turbulence are best distinguished by \shape, waves and balanced flows are better distinguished by \sparsity.
IGWs are less sparse for $j_2 \geq 3$ and smoother in shape for $j_{\{1, 2\}} \lesssim 4$, trending towards the reference value $\shape = 1$ for $j_2 > 4$.
In the time series of \sparsity\ and \shape\ at the \numrange{4}{8} grid-point scale ($j_1 = 2$, $j_2 = 3$), IGW and balanced simulations start similarly, due to their random linear-mode initialization, but diverge as nonlinear turbulence develops in the balanced case (Fig.~\ref{fig:fig3}c).
The balanced flow shows a more pronounced change in time with an initial rapid increase in \sparsity\ that plateaus around $\sparsity \approx 0.16$.
The \shape\ steadily increases, indicating a progressive sharpening of the small scales.
IGWs are much less variable with nearly constant $\sparsity \approx 0.12$ and $\shape \approx 1$.
The slow evolution due to weak damping and weak wave--wave interactions smoothens the waves over time ($\shape < 1$).
At larger scales, \numrange{32}{64} grid points ($j_1 = 5$, $j_2 = 6$), sparsity in the balanced flow increases gradually, while shape oscillates between smooth ($\shape < 1.0$) and sharp ($\shape > 1.0$), reflecting repeated vortex mergers forming larger vortices.
This aligns with expectations for an inverse energy cascade: large scales become sparser, and small scales sharpen as filaments form.
These large scales are only weakly excited in the IGW simulations, so their time series are not shown.

\section{Realistic Flows in the Upper Ocean} \label{sec:sec5} 

\begin{figure}
  \centering
  \includegraphics[width=1\textwidth]{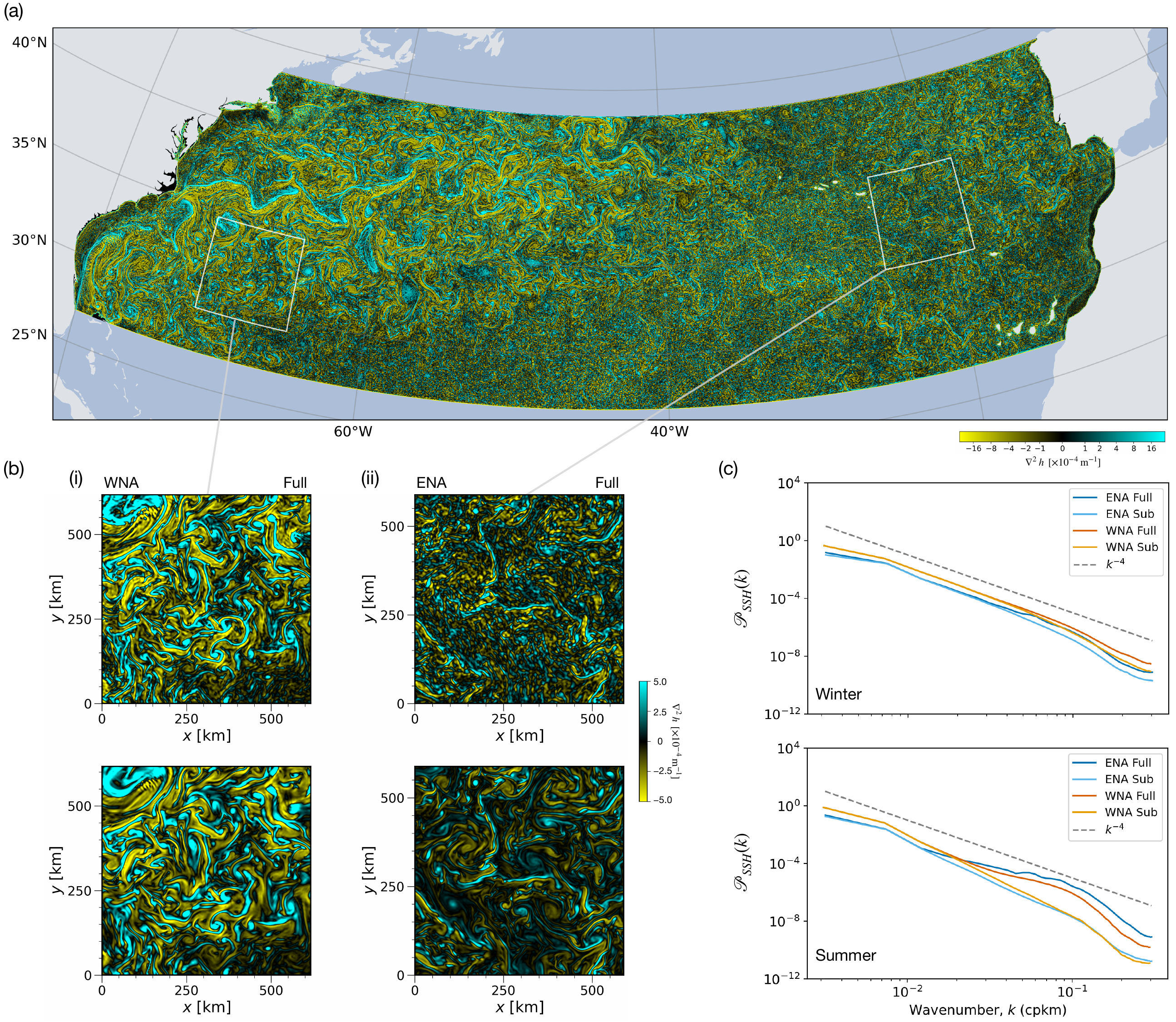}
  \caption{
    Waves and balanced flows in a realistic simulation of the North Atlantic Ocean.
    (a)~Snapshot of $\nabla^2 h$, where $h$ is SSH, on March~3 (using an arsinh-scaled color map).
    (b)~Magnified regions of interest in the western and eastern North Atlantic, where the fields at small scales are dominated by submesoscale turbulence and waves respectively.
    (i, ii)~show the full fields; (iii, iv) show the subinertial fields obtained by temporal filtering.
    (c)~SSH isotropic power spectral density (in units ${\rm m}^2 /{\rm cpkm}$) for the full and subinertial fields (top and bottom panels, respectively) averaged over 90~days in summer (6/6 to 9/04) and winter (12/06 to 3/06).
    Dashed lines indicate a $k^{-4}$ spectral slope.
    }%
\label{fig:fig4}
\end{figure}

In the real ocean, the meso- and submesoscale SSH field is a superposition of signals arising from balanced turbulence and internal waves, both internal tides and the broadband continuum.
To understand whether the ST can disentangle these, we apply it to SSH fields from a realistic North Atlantic simulation with an $\sim$\qty{2}{\kilo\meter} grid spacing \citep[][Fig.~\ref{fig:fig4}a]{Sinha_2023}.
We analyze the SSH of two dynamically distinct approximately $\qty{600}{\kilo\meter} \times \qty{600}{\kilo\meter}$ regions, located in the eastern North Atlantic (ENA) and western North Atlantic (WNA, Fig.~\ref{fig:fig4}b).
The WNA has strong balanced turbulence, with intense mesoscale activity driven by the Gulf Stream and its extension, generating eddies, fronts, and filaments.
The ENA is more quiescent with weaker mesoscale activity and a greater relative contribution of internal waves.
Despite the regions' clear dynamical differences, power spectral analysis struggles to distinguish between the ENA and WNA (Fig.~\ref{fig:fig4}c).
During winter, both regions have similar SSH spectra with a $k^{-4}$ power law, despite more prominent waves in the ENA (Fig.~\ref{fig:fig4}b).
In summer, the increased presence of internal waves flattens both SSH spectra in the \qtyrange{10}{100}{\kilo\meter} range, clarifying the transition from balanced to wave-dominated signals in the spectrum.
Differences in the dynamics of the balanced meso- and submesoscale turbulence dominating at larger scales, however, cannot be assessed using spectral analysis.

In models, frequency filtering can separate SSH fields into waves and balanced components, providing a test of the ST's ability to distinguish them for satellite altimetry fields, where limited temporal resolution precludes such separation.
We isolate balanced signals using a 4th-order Eulerian Butterworth filter, applied to hourly SSH snapshots with a cutoff frequency at the local Coriolis frequency.
This removes high-frequency wave motions leaving only the subinertial component associated with balanced dynamics.\footnote{Lagrangian filtering would be preferable but the Eulerian filter is sufficient for the ST analysis.}
Differences between the full and subinertial SSH fields indicates prominent wave signals, while agreement indicates the dominance of balanced motions.
We apply the ST to both the ``full'' $\nabla^2 h$ fields and the ``subinertial'' filtered $\nabla^2 h$ fields, where $h$ is SSH.

\begin{figure*}
  \centering
  \includegraphics[width=1\textwidth]{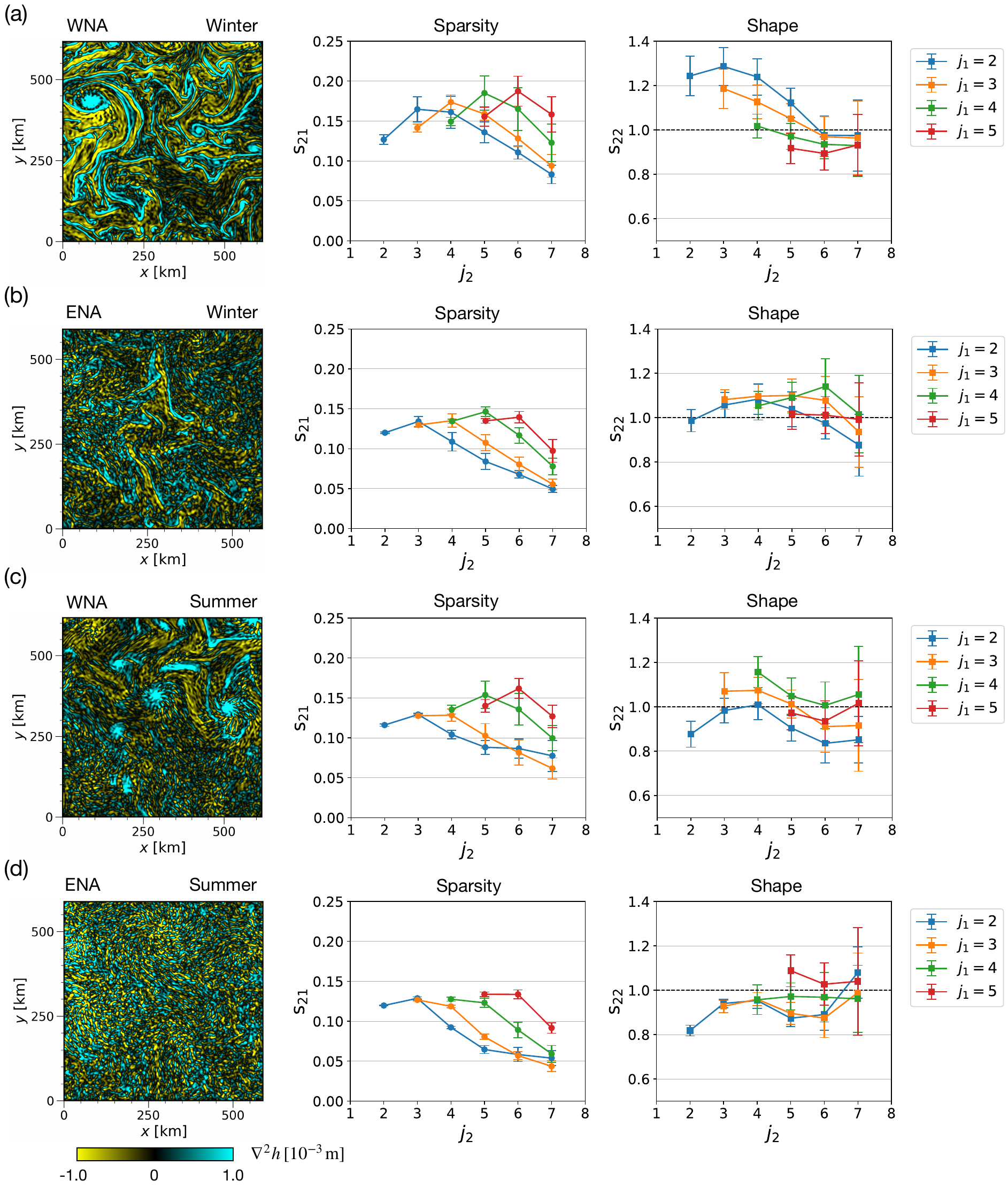}
  \caption{
    Characterizing the shape and sparsity of realistic fields.
    Shown are snapshots of $\nabla^2 h$ in the ENA and WNA regions with axes in units of grid points and 90-day averages of the shape and sparsity coefficients for $j_1 = \numrange{2}{5}$ and $j_2 = \numrange{2}{7}$, characterizing structures with spatial scales between \qtylist{4;128}{\kilo\meter}.
    Error bars show standard deviations.
  }
\label{fig:fig5}
\end{figure*}

The ST distinguishes the ENA and WNA regions by their \shape\ and \sparsity\ values (Fig.~\ref{fig:fig5}).
For both regions, we inspect \shape\ and \sparsity\ averaged over 90 daily snapshots in the summer and winter for spatial scales of \qtyrange{4}{32}{\kilo\meter} for $j_1$ and \qtyrange{4}{128}{\kilo\meter} for $j_2$.
The ENA has overall lower \shape\ and \sparsity\ values compared to the WNA, consistent with enhanced wave activity in the ENA and dominant eddies and fronts in the WNA.
Seasonal differences are also clear in both regions, with larger \sparsity\ and \shape\ values in winter than in summer indicating waves dominate the SSH field in summer, while fronts and eddies contribute more in winter for the WNA.
This seasonality aligns with expectations: submesoscale fronts and eddies are energized in winter by a deepened mixed layer triggering mixed-layer instability (MLI) \citep{mensa_2013, Sasakietal_2014, callies_2015, callies_2016}, whereas the wave signature in SSH increases in summer with the enhanced near-surface stratification \citep{rocha_2016, lahaye_2019}.

\begin{figure*}
  \centering
  \includegraphics[width=1.0\textwidth]{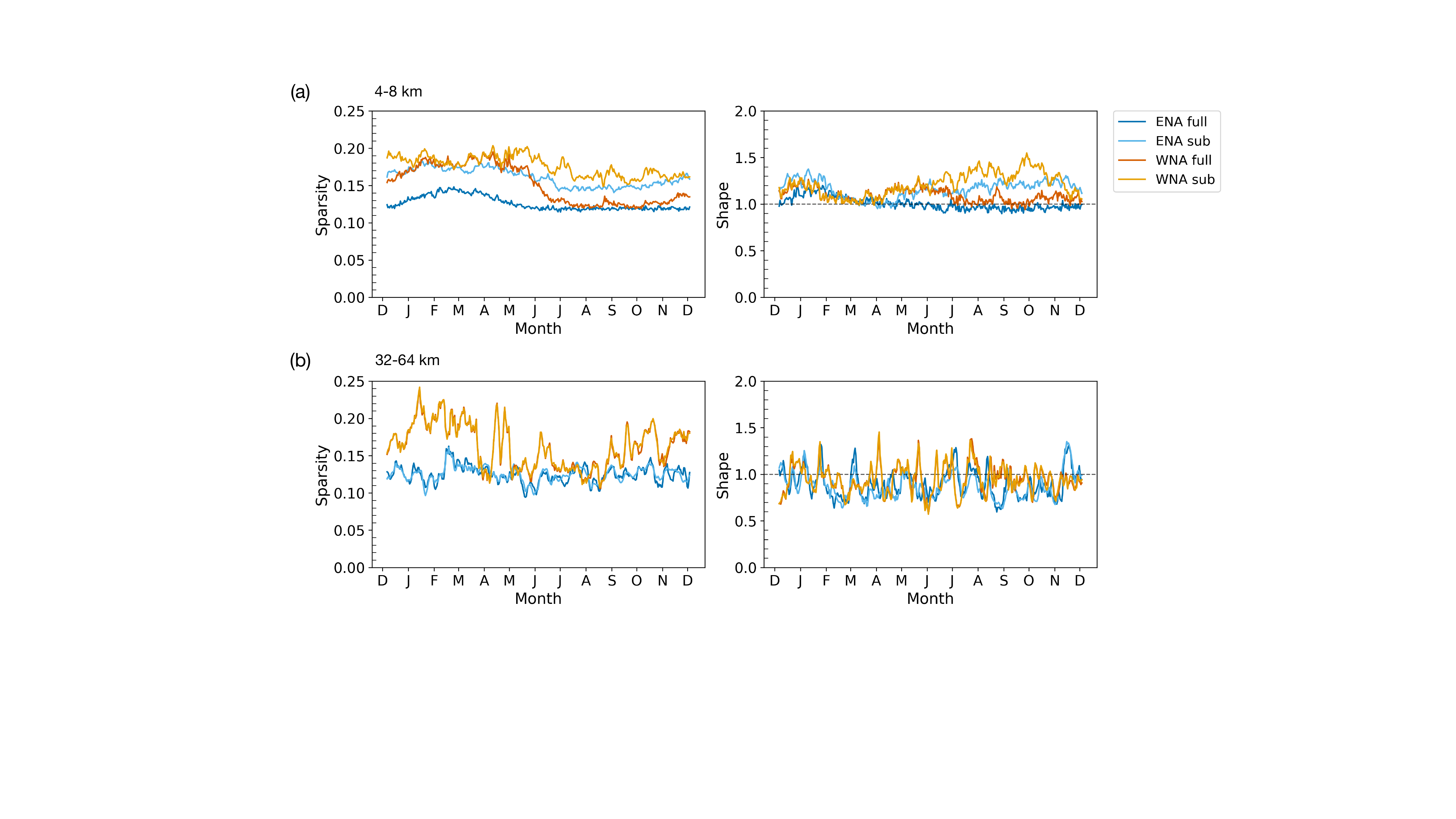}
  \caption{
    Seasonal evolution of the shape and sparsity in the ENA and WNA regions.
    Shown are time series of sparsity~\sparsity\ and shape~\shape\ from daily snapshots of $\nabla^2 h$.
    Values are shown for both the full and subinertial fields for spatial scales of (a)~\qtyrange{4}{8}{\kilo\meter} ($j_1 = 2$, $j_2 = 3$) and (b)~\qtyrange{32}{64}{\kilo\meter} ($j_1 = 5$, $j_2 = 6$).
  } 
\label{fig:fig6}
\end{figure*}

A clear seasonal progression appears in the time series of daily \sparsity\ and \shape\ for the ENA and WNA regions, reflecting shifting contributions of waves and balanced flows (Fig.~\ref{fig:fig6}).
At \qtyrange{4}{8}{\kilo\meter}, the ENA full fields maintain lower \sparsity\ year-round than the ENA subinertial and WNA fields, indicating a consistently more wave-dominated SSH field in the ENA (Fig.~\ref{fig:fig6}a).
In winter, \sparsity\ and \shape\ increase in both regions, and the close match between the full and subinertial fields in the WNA suggests minimal wave contribution to SSH.
A sharp drop in sparsity to $\sparsity \approx 0.12$ and shape to $\shape \approx 1.0$ marks the transition to dominant wave activity around May in the ENA and July in the WNA.
Visually inspecting the $\nabla^2h$ fields confirms that the WNA's delayed drop in \sparsity\ is due to its more energetic submesoscale fronts and eddies, which take longer to decay before waves take over.
Once wave-dominated, the SSH fields maintain nearly constant \sparsity\ throughout the summer, persisting until December.
At \qtyrange{32}{64}{\kilo\meter}, the shape and sparsity time series of the full and subinertial fields overlap almost exactly (Fig.~\ref{fig:fig6}b).
This indicates that these larger-scale dynamics are dominantly balanced, even when internal waves are excited in the simulation.
This close agreement suggests that the ST can distinguish between wave-dominated and balanced flow regimes across different spatial scales.
The consistently greater sparsity in the WNA compared to the ENA at these scales also points to differences in the mesoscale balanced flow between the regions, beyond just their energy levels.

\section{Discussion} 

As demonstrated through a series of idealized and realistic numerical simulations, the ST can effectively characterize the structure of upper-ocean flow fields.
The ST provides information beyond that captured by the power spectrum, quantifying the geometry of flow features---e.g., their shape and sparsity---across scales.
This geometry is sufficiently different for internal waves and different flavors of balanced meso- and submesoscale turbulence so can be picked up robustly by the ST.
We expect that an application of the ST to data from the SWOT mission, for example, would provide important insight into the dynamics underlying the observed SSH snapshots.
Moreover, the ST could be used to compare the representation of submesoscale turbulence in different models, realistic and idealized.
Comparing the ST between models and observations could provide a more stringent test of the models' ability to capture the underlying dynamics than is possible with spectral analysis.
Models must capture not only the correct power at each scale but also the geometry of jets, eddies, fronts, filaments, and waves to compare favorably with observations.

While the ST is already informative, theoretical advances could further improve its utility.
For example, identifying wavelets that satisfy the Littlewood–Paley criterion---i.e., whose squared Fourier amplitudes sum to a constant across wavenumbers such that the sum of the squares of the scattering coefficients is equal to the signal variance \citep{Mallat_2012}---would yield a more interpretable, energy-preserving representation of the fields.
Though not essential for comparative analyses (as in the present work), this property would clarify how much of the signal is captured by low-order scattering coefficients and when higher orders are needed \citep{cheng_2021}.
Finally, establishing a rigorous theory for ST-derived metrics---analogous to that for spectral slopes---would deepen our understanding of the physical meaning of scattering coefficients and their relationship to fundamental turbulence properties.

\section*{Open Research}

The data and software used in this research are publicly available and can be accessed as follows:
\begin{itemize}
  \item The scattering transform codes:  \url{https://github.com/SihaoCheng/scattering_transform} \citep{cheng_2021}; \url{https://github.com/kymatio/kymatio} \citep{Andreuxetal_2020}
  \item Dedalus,  \url{https://github.com/DedalusProject/dedalus} \citep{Dedalus}
\end{itemize}

\acknowledgments 
This work was supported by NASA grant 80NSSC23K0345 and a Simons 
Foundation Pivot Fellowship.
This work used high performance computing at the San Diego 
Super Computing Centre \citep{SDSC2022} through allocation PHY-230189 from 
the Advanced Cyberinfrastructure Coordination Ecosystem: 
Services \& Support (ACCESS) program, which is supported 
by National Science Foundation grants \#2138259, \#2138286, 
\#2138307, \#2137603, and \#2138296.
This work also used high-performance computing awarded to JWS by 
the Google Cloud Research Credits program GCP19980904.
The authors thank Sihao Cheng, Marco Gatti, Bhuvnesh Jain, Patrice Klein, Brice Ménard, Dimitris Menemenlis and Andrew Thompson for helpful discussions.

\section{Supplementary Materials}

\subsection{Scattering Transform Analysis}

For our ST calculations, we use the \texttt{scattering} package by \citet{cheng_2021}.
We have also tested the \texttt{kymatio} package by \citet{Andreuxetal_2020} and found agreement between the two.
For the ST setup, we use $J = 8$ and $L = 4$ for the presented flow fields at $(N_x, N_y) = (256, 256)$ horizontal resolution. 
We have verified that our results are unchanged up to $(N_x, N_y) = (1024, 1024)$ resolution with $J$ adjusted to $J = \log_2 N_x$. Flow fields are obtained from a hierarchy of models described below.

\subsection{Turbulence Models}

The turbulence simulations presented are performed with a doubly periodic spectral model with a resolution of $(N_x, N_y) = (256, 256)$. 
The governing equations of the model permit multiple flavors of turbulence and cascade directions to be simulated via the material advection of a scalar field~$\xi$ and an $\alpha$-dependent relation between $\xi$ and the streamfunction~$\psi$:
\begin{equation}
  \frac{\partial \xi}{\partial t} + \J(\psi, \xi) = \mathcal{F} - \mathcal{D}, \qquad \xi = \nabla ^{\alpha} \psi,
\label{eq:eq_PV}
\end{equation}
where $\J(A, B) = \partial_x A \, \partial_y B - \partial_y A \, \partial_x B$ is the Jacobian operator.
The $\mathcal{F} = \mathcal{F}(k)$ denotes wavenumber dependent forcing centered on wavenumber~$k_f$, and $\mathcal{D} = \mathcal{D}(k)$ is a combination of linear drag and an 8$^{\rm th}$-order hyper-viscosity.
The flavor of the turbulence and the cascade direction are controlled by $\alpha$ and $k_f$, respectively \citep{Pierrehumbert_1994, smith_2002}.
We set $\alpha = 2$ and $\alpha = 1$ with $k_f$ set to high and low wavenumbers to simulate 2D and SQG turbulence with inverse and forward cascades, respectively.
With $\alpha = 2$ and $\alpha = 1$, $\xi$ in \eqref{eq:eq_PV} is vorticity and surface buoyancy, respectively \citep{held_1995}.

The turbulence simulations are initialized with two-dimensional Gaussian random fields with spectral kinetic energy distribution proportional to $\mathcal{E}_k \propto k^{-5/3}$ where $k$ is the magnitude of the horizontal wavenumbers:
\begin{equation}
  |\hat{\psi}(k)|^2 \propto \frac{k^{-2}}{\left[1 + \left(\frac{k}{k_0}\right)^2\right]^{\frac{1}{2}(s + 1)}},
\end{equation}
where $k_0 = 5$ sets the location of the spectral peak and $s = 5/3$ sets the slope for $k \gg k_0$. The simulations then evolve under the applied $\mathcal{F}$ and $\mathcal{D}$ until they equilibrate, judged by inspecting time series of kinetic energy.

\subsection{Shallow-Water Simulations}

For the balanced vs.\ gravity wave simulations, we use a spectral model of the rotating shallow-water equation (SWE) implemented in \texttt{Dedalus} \citep{Dedalus}. 
We use the modification of the SWEs proposed by \citet{buhler_shallow-water_1998} to avoid nonlinear steepening of gravity waves.
The model uses $3/2$ dealiasing, a 4th-order Runge--Kutta time integration scheme, and doubly periodic boundary conditions.
The domain of the simulations presented has spatial resolution of $(N_x, N_y) = (256, 256)$ with spatial extent $(L_x, L_y) = (-10, 10) \, L_d$ where $L_d = \sqrt{g H}/{f}$ is the Rossby radius of deformation.  
The SWEs are solved in non-dimensional form with Rossby and Froude numbers $\mathrm{Ro} = 0.1$ and $\mathrm{Fr} = 0.01$, and with initial conditions derived from geostrophic balanced or inertia-gravity wave solutions to the linearized SWEs \citep{Vallis2017}.
The initial velocity and height fields are Guassian random fields with kinetic energy spectra representative of the Garrett--Munk spectrum \citep{Garrett&Munk_1972}, peaking at low wavenumbers, having a $k^{-2}$ power law below, and being tapered exponentially at high wavneumbers. 
The SWE model evolves the equations freely in time with no additional forcing or dissipation besides an $8^{\rm th}$-order hyper-viscosity.

\subsection{North Atlantic Simulation}

Our simulation of the North Atlantic uses the setup of the MITgcm described in \citet{Sinha_2023}.
We run one year of this model starting on model date 06-12-2025 from an output file produced from the simulation described by \citet{Sinha_2023}.
We save hourly snapshots of the surface fields.
One difference in setup between the simulations in the present work and \citet{Sinha_2023} is that our open boundary conditions vary monthly, whereas the original setup used a constant annual mean.

\bibliography{biblio.bib}

\end{document}